\documentclass[11pt,a4paper]{article}
\usepackage{jcappub}

\def\tsutsui{$E_{\rm p}$--$T_{\rm L}$--$L_{\rm p}$ }

\title{Gamma-Ray Bursts are  precise distance indicators similar to \\ Type Ia Supernovae?}
                                            \author[a]{Ryo Tsutsui,}
                                            \author[b]{Takashi Nakamura,}
                                            \author[c]{Daisuke Yonetoku,}
				      \author[d]{Keitaro Takahashi,}
				         \author[c]{and Yoshiyuki Morihara}
                                            \affiliation[a]{Research Center for the Early Universe, School of Science, University of Tokyo, Bunkyo-ku, 
				         Tokyo 113-0033, Japan}
                                            \affiliation[b]{Department of Physics, Kyoto University, Kyoto 606-8502, Japan} 
				      \affiliation[c]{Department of Physics, Kanazawa University, Kakuma, Kanazawa,
				      Ishikawa 920-1192, Japan}
				      \affiliation[d]{Faculty of Science, Kumamoto University, Kurokami, Kumamoto, 860-8555, Japan}
				       \emailAdd{tsutsui@resceu.s.u-tokyo.ac.jp}
                                            \emailAdd{takashi@tap.scphys.kyoto-u.ac.jp}
                                            \emailAdd{yonetoku@astro.s.kanazawa-u.ac.jp}
				       \emailAdd{keitaro@sci.kumamoto-u.ac.jp}
				       \emailAdd{morihara@astro.s.kanazawa-u.ac.jp}
                                            \abstract{We estimate the distance modulus to long gamma-ray bursts (LGRBs) using the  Type I 
Fundamental Plane, a correlation between the spectral peak energy $E_{\rm p}$, 
the peak luminosity $L_{\rm p}$, and the luminosity time $T_{\rm L}$ ($\equiv E_{\rm iso}/L_{\rm p}$
 where $E_{\rm iso}$ is isotropic energy) for small Absolute Deviation from Constant Luminosity(ADCL).
The Type I Fundamental Plane of LGRBs is calibrated using 8 LGRBs with redshift $z<1.4$. 
To avoid any assumption on the cosmological model, we use the distance modulus of 557 
Type Ia supernovae (SNeIa) from the Union 2 sample. 
This calibrated Type I Fundamental Plane is used to measure the distance moduli to 
9 high-redshift LGRBs with the mean error $\bar \sigma_{\mu}=0.31$, which is comparable with 
that of SNe Ia  $\bar \sigma_{\mu}=0.26$ where $\mu$ stands for the distance modulus.
The Type I Fundamental Plane is so tight that our distance moduli 
have very small uncertainties. 
From those distance moduli, we obtained the constraint 
$\Omega_{\rm M}=0.22\pm0.04$ for flat $\Lambda$CDM universe.  
Adding 9 LGRBs distance moduli ($z>1.4$) to 557 SNeIa distance moduli ($z<1.4$) significantly 
improves the constraint for non-flat $\Lambda$CDM universe from   
($\Omega_{\rm M}, \Omega_{\rm \Lambda}$)=($0.29\pm0.10$, $0.76\pm0.13$ )  for SNeIa only to
($\Omega_{\rm M}, \Omega_{\rm \Lambda}$)=($0.23\pm0.06$, $0.68\pm0.08$)   for SNeIa and 9 LGRBs.
}
                                            \arxivnumber{1205.2954}

\begin{document}

\maketitle

\section{Introduction}
Since the discoveries of possible distance indicators of long gamma-ray bursts (LGRBs) 
\citep{Fenimore:2000,Norris:2000,Amati:2002,Yonetoku:2004,Ghirlanda:2004a, Liang:2005,Firmani:2006}, 
many attempts to measure the distance up to the high redshift universe ($z > 1.7$) which
cannot be probed by Type Ia supernovae have been done to test cosmological models
\citep{Bloom:2003,Schaefer:2003,Ghirlanda:2004b, Schaefer:2007, Kodama:2008,Liang:2008,
Amati:2008,Tsutsui:2009a,Cardone:2009,Tsutsui:2009b,Wei:2009,Wei:2010,Diaferio:2011}. These attempts are very important
since no other distance indicators have been known in such high redshift universe.

However the large statistical and systematic errors of the distance indicators
restricted the precision of distance measurements by LGRBs. 
To estimate distances of LGRBs more precisely, there are two directions : to discover more precise distance
indicator \citep{Ghirlanda:2004b,Firmani:2006b,Liang:2005} and/or to take weighted average of distances
derived by many distance indicators \citep{Schaefer:2007,Liang:2008,Cardone:2009}. 
These studies, however, are on the verge of many difficulties, such as missing or multiple jet breaks
in {\it Swift era} \citep{Ghirlanda:2007,Sato:2007} or increasing scatters of distance indicators
\citep{Butler:2007,Rossi:2008,Collazzi:2008}. To make LGRBs to be more precise and reliable standard candles,
we must know the origins of the dispersion of distance indicators to control them carefully. 

In our previous studies, we have investigated origins of dispersion of the spectral - brightness correlations
and found possible origins of systematic errors \citep{Yonetoku:2010,Tsutsui:2010}: 
(i) spectral peak energies ($E_{\rm p}$) estimated by fitting the spectra with cut-off power-law model
with three free parameters  tend to be overestimated compared with those obtained by using the Band model
with four free parameters \citep{Band:1993},
(ii) peak luminosities ($L_{\rm p}$) should be estimated for a fixed time scale in the individual rest frame rather than the observer frame, 
(iii) contamination of short GRBs or other unknown populations.
Removing the systematic errors by (i) and (iii), and correcting (ii), we found that the Fundamental Plane of LGRBs,
that is, a correlation between the spectral peak energy $E_{\rm p}$, the rest frame 2.752-second peak luminosity $L_{\rm p}$,
and the luminosity time $T_{\rm L}$ ($\equiv E_{\rm iso}/L_{\rm p}$ where $E_{\rm iso}$ is the isotropic energy),
is much tighter than the original $E_{\rm p}-T_{\rm L}-L_{\rm p}$ correlation by R. Tsutsui et al. \citep{Tsutsui:2009b}
\citep[See also ] []{Tsutsui:2011}.

\begin{figure}[ht]
\begin{center}
  \includegraphics[width=80mm]{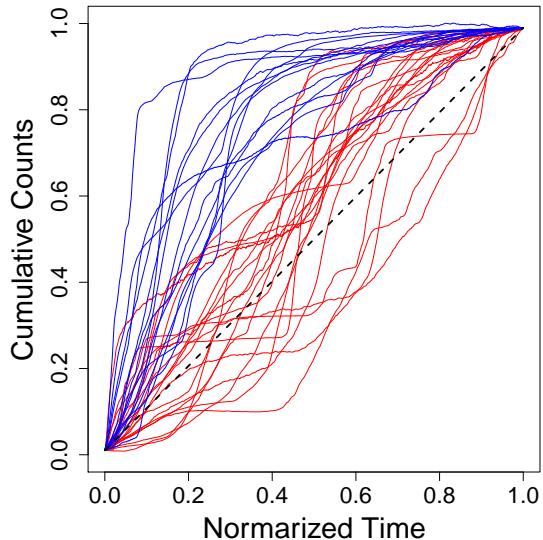}
\vspace{0pt}
\caption{The cumulative light curves of  LGRBs from  Tsutsui, et al. \citep{Tsutsui:2012a}.
The red solid  lines indicate small-$ADCL$ events  while blue solid  lines do
 large-$ADCL$ events. The black dashed line indicates the line of the virtual source with the constant luminosity. 
We clearly see that red lines with small ADCL ($<0.17$) and blue solid lines with large ADCL($>0.17$) form different groups.
}
\label{fig:cumulative}
\end{center}
\end{figure}

Recently, we found that dividing LGRBs into two groups according to the {\it absolute deviation from their constant luminosity}
($ADCL$) significantly improves the Fundamental Plane of LGRBs \citep{Tsutsui:2012a}.
$ADCL$ is of critical importance in this work so let us summarize the main results below.
$ADCL$ is a quantity which characterizes the overall shape of light curve of a GRB and defined as follows:
\begin{equation}
\label{eq:ADCL}
ADCL = \sum_{i=1}^{N_{\rm bin}} \frac{|C_{i}^{\rm norm}-0.01-0.98\times t_i^{\rm
 norm}|}{N_{\rm bin}},
\end{equation}
where $C_{i}^{\rm norm}$ is  normalized cumulative counts and $ t_i^{\rm norm}$ is
the normalized time by $T_{\rm 98}$ in which  the cumulative counts increase from 0.01 to 0.99. 
Most of small-$ADCL(<0.17)$ events consist of many pulses with almost the same brightness,
while most of large-$ADCL(>0.17)$ events consist of a bright main pulse and the weak long tail like FRED (Fast Rise and Exponential Decay).
In Figure. \ref{fig:cumulative}, we show the cumulative light curves of  LGRBs from  Tsutsui, et al. \citep{Tsutsui:2012a}.
The time  and the cumulative counts are normalized by $T_{\rm 98}$ and the total counts, respectively.
The red solid  lines indicate small-$ADCL$ events  while blue solid  lines do large-$ADCL$ events. 
The black dashed line indicates the line of the virtual source with a constant luminosity.
We clearly see that red lines with small ADCL ($<0.17$) and blue solid lines with large ADCL($>0.17$) form different groups in this
figure so that our classification scheme for subclasses of LGRB would be reasonable.

\begin{figure}[ht]
\begin{center}
\begin{tabular}{cc}
\includegraphics[width=80mm]{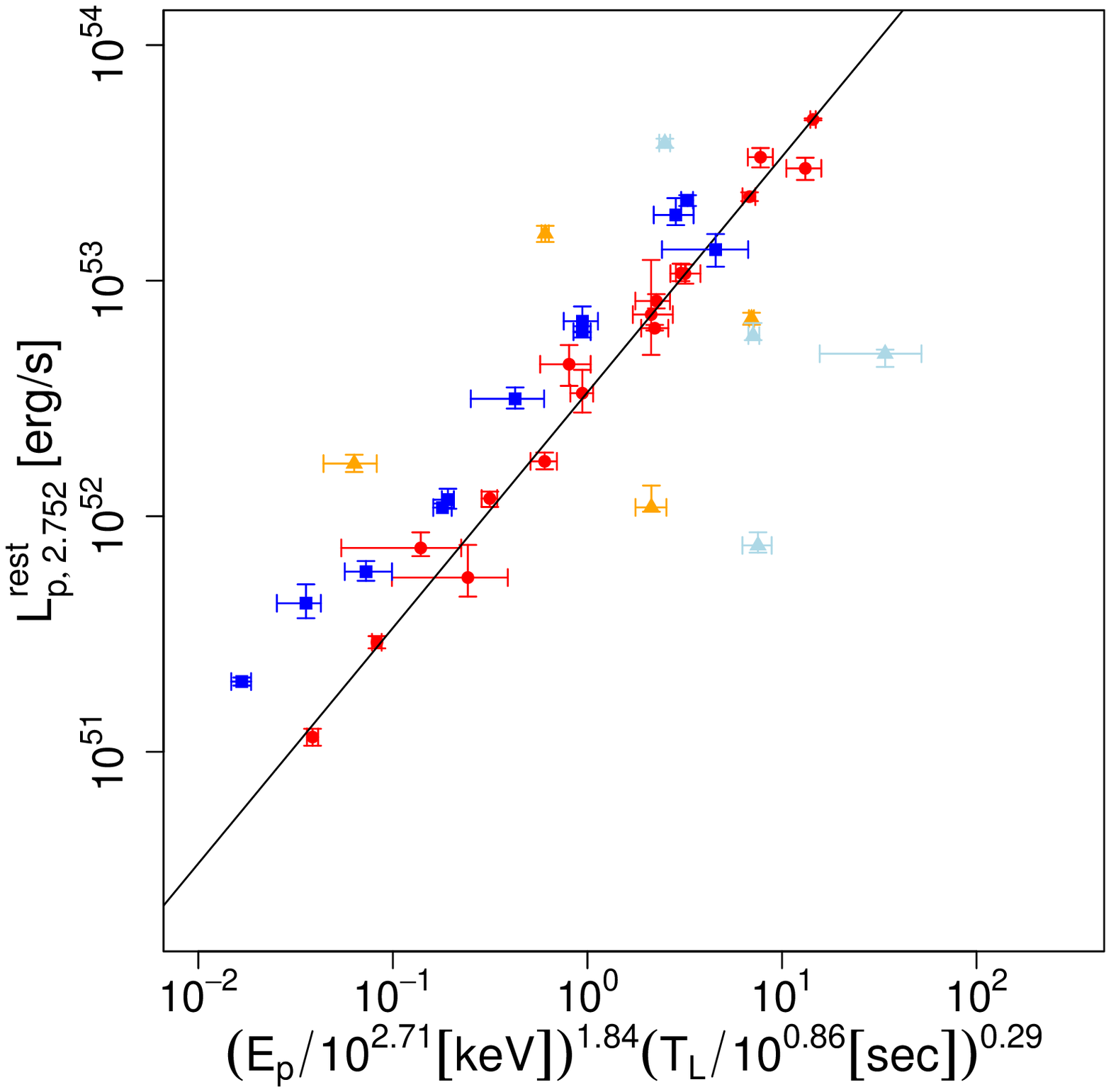}
\includegraphics[width=80mm]{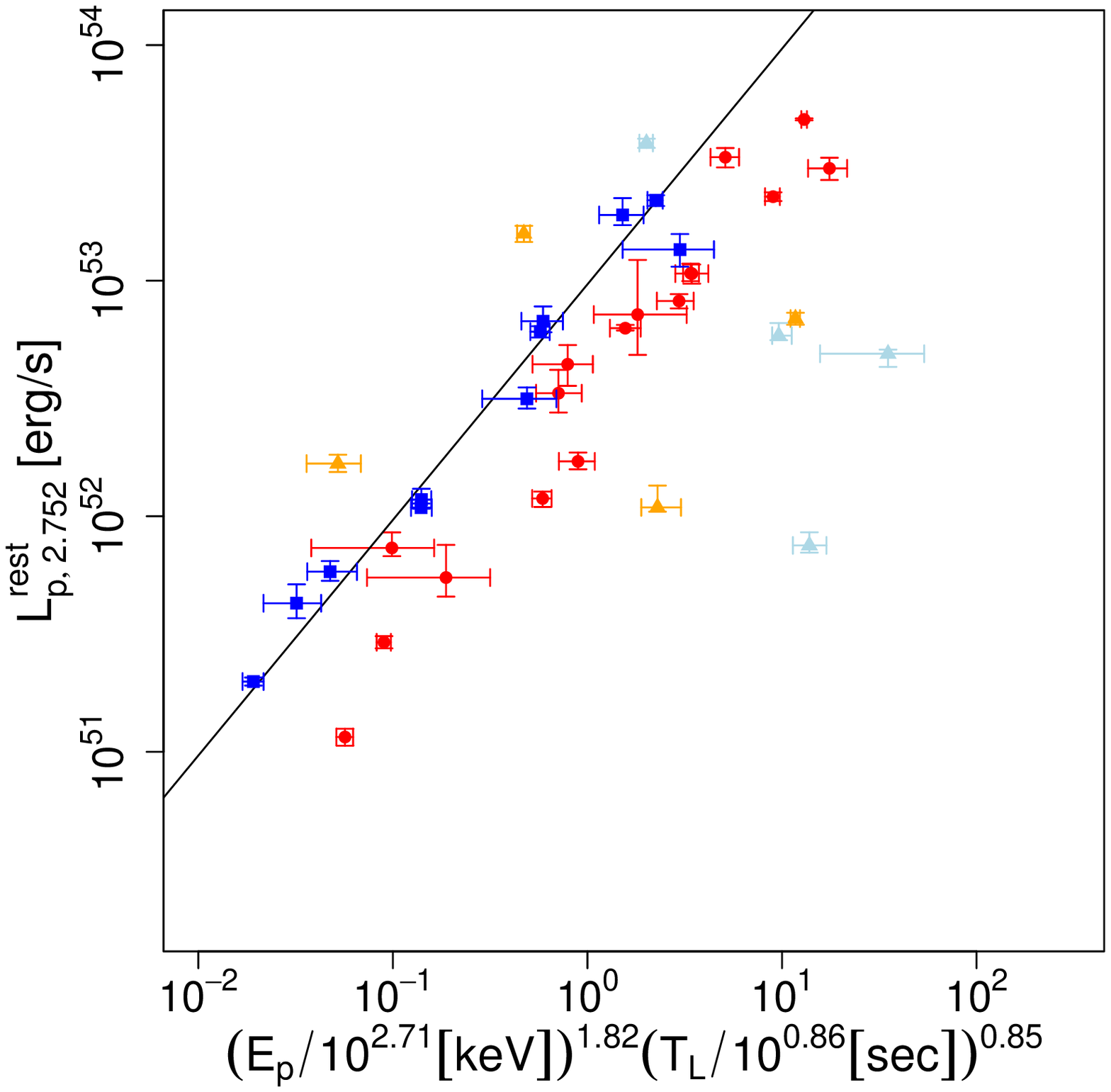}
\end{tabular}
\vspace{0pt}
\caption{The \tsutsui diagram both for the Type I (left) and Type II (right) Fundamental Planes for given cosmological parameters
 from  Tsutsui, et al. \citep{Tsutsui:2012a}. Small-$ADCL$ events are marked with red filled circles and large-$ADCL$ events 
with blue squares.  Except for  outliers ( orange triangles), 
most of small-$ADCL$ events are on the Type I Fundamental Plane (=the solid line in the left figure) 
within their error bars {\it without  any systematic errors}, 
while most of large-$ADCL$ events are on the Type II Fundamental Plane (=the solid  line in the right figure)
except for outliers (light blue triangles). (see \citep{Tsutsui:2012a} for the definition of outliers. )
In the  left  figure, blue squares which belong to large ADCL events are above the
solid line (=the Type I Fundamental Plane) while in the right  figure, red filled circles which belong to small ADCL events are
below the solid line (=the Type II Fundamental Plane) so that the existence of two Fundamental Planes can be recognized.
The existence of such two subclasses   is very  
 similar to that of   two   Period - Luminosity relations of  Cepheid variables.
}
\label{fig:FP}
\end{center}
\end{figure}

In Tsutsui, et al. \citep{Tsutsui:2012a}, we found that the best-fit function for small ADCL($<0.17$) events is given by
\begin{eqnarray}
\label{eq:FP1}
L_{\rm p}=10^{52.53\pm 0.01}{\rm erg s^{-1}}&&
\left(\frac{E_{\rm p}}{10^{2.71} 
{\rm keV}}\right)^{1.84\pm 0.03} \nonumber \\
&& \times \left(\frac{T_{\rm L}}{10^{0.86}{\rm sec}}\right)^{0.29\pm0.08},
\end{eqnarray}
with $\chi^2_{\nu}/d.o.f=10.93/14$ and $\sigma_{\rm int}=0$,
while that for large ADCL($>0.17$) events is given by
\begin{eqnarray}
\label{eq:FP2}
L_{\rm p}=10^{52.98\pm 0.080}{\rm erg s^{-1}}&&
\left(\frac{E_{\rm p}}{10^{2.71}{\rm keV}}\right)^{1.82\pm
0.093}\nonumber \\ 
& &\times \left(\frac{T_{\rm L}}{10^{0.86}{\rm sec}}\right)^{0.85\pm0.26},
\end{eqnarray}
with $\chi^2_{\nu}/d.o.f=7.58/8$ and $\sigma_{\rm int}=0$.
Since both best-fit functions  separately form the plane in $(\log E_p, \log T_L , \log L_p)$ space,  
 we call the best fit function  for small-$ADCL$ (large-$ADCL$)  as Type I (Type II) Fundamental Plane, respectively.
Figure. \ref{fig:FP} shows the Type I (left) and Type II (right) Fundamental Planes
 for given cosmological parameters from Tsutsui, et al. \citep{Tsutsui:2012a}.
 In both figures, Small-$ADCL$ events are marked with red filled circles   
 and large-$ADCL$ events are with blue squares.
 Orange and light blue squares indicate outliers from the Type I and Type II Fundamental Planes, respectively.
(see \citep{Tsutsui:2012a} for the definition of outliers. )
 The left (right) of Figure. \ref{fig:FP} is the plane perpendicular to the Type I (Type II) Fundamental Plane 
so that the Type I (Type II) Fundamental Plane is expressed by the solid line in each figure. 
 In the  left of Figure. \ref{fig:FP}, blue triangles which belong to the large ADCL events are above the
solid line (=the Type I Fundamental Plane) while in the right of Figure. \ref{fig:FP}, red filled circles which belong to the small ADCL events are
below the solid line (=the Type II Fundamental Plane) so that the existence of two Fundamental Planes can be recognized.
The existence of such two subclasses   is very  
 similar to that of   two   Period - Luminosity relations of  Cepheid variables \citep{Baade:1956}.

In this paper using  21 small-$ADCL$ LGRBs from the 36 gold sample of  Tsutsui et al. \citep{Tsutsui:2012a}, 
we try to measure distance modulus up to $z=3.57$ and determine the cosmological parameters. 
To avoid any assumption on cosmological models, we use the distance moduli of 557 
Type Ia supernovae (SNeIa) from the Union 2 sample to calibrate the Type I Fundamental Plane of LGRBs 
by the 10 low-z events in $0.168<z<1.30$. This is a big difference from our previous paper \citep{Tsutsui:2012a}
where cosmological parameters were assumed to derive the relation.
Applying the Type I Fundamental Plane calibrated at low redshifts without cosmological parameters to high-z GRBs,
we obtain the relation between the redshift and the distance modulus $\mu$ of 11 high-z events in $1.54<z<3.57$. 
From these distance moduli, we will obtain the constraint on  $\Omega_{\rm M}=0.22\pm0.04$ for flat $\Lambda$ CDM universe.
In this paper we do not use the Type II Fundamental Plane as a distance indicator to
constrain the cosmological parameters due to the lack of enough  number of large ADCL events.
 Throughout the paper, we fix the Hubble parameter as $H_0 = 70$ km s$^{-1}$ Mpc $^{-1}$.

\section{Calibration}

\subsection{local regression}
First, we calibrate the Type I Fundamental Plane of LGRBs using low-z events whose distance moduli
can be obtained by those of Type Ia supernovae. Actually, to estimate the distance moduli of low-z GRBs,
we use a local regression method first developed by W. S. Cleveland. \citep{Cleveland:1979} and
W. S. Cleveland and S. J. Devlin. \citep{Cleveland:1988}, and first applied for GRB cosmology
by V. F. Cardone et al. \citep{Cardone:2009}. This method does not need a global shape function of
redshift-distance modulus relation and is one of interpolation procedures which use only the data
near points and fit them with a low-degree polynomial. Although the number of data and the degree
of polynomial are arbitrary variables, it does not depend on a shape of a global function.

Applying this method, we obtain the distance moduli of 10 low-redshift LGRBs with mean error
$\sigma_{\mu}=0.065$ where $\mu$ stands for the distance modulus.

\subsection{calibration}
Having estimated the distance moduli of 10 low-redshift GRBs in a model independent way,
we can calibrate the Type I Fundamental Plane of GRBs.
Let us assume a linear correlation
between $E_{\rm p}$, $T_{\rm L}$ and $L_{\rm p}$ in logarithmic scale as,
\begin{equation}
\log L_{\rm p} (E_{\rm p},T_{\rm L})
= A + B \log \left(E_{\rm p} / {\bar E_{\rm p}}\right)
  + C \log \left(T_{\rm L} / {\bar T_{\rm L}}\right),
\label{eq:fp}
\end{equation}
where $A$, $B$ and $C$ are the parameters of the model and 
$\bar E_{\rm p}$ and $\bar T_{\rm L}$ are logarithmic mean of $E_{\rm p}$ and $L_{\rm p}$ with 
$10^{2.43}$ keV and $10^{0.97}$ sec, respectively.
From equation (\ref{eq:fp}), we obtain $\mu(E_{\rm p},T_{\rm L}, F_{\rm p})$ given by,
\begin{eqnarray}
 \mu (E_{\rm p},T_{\rm L},F_{\rm p})
=& & 2.5 \log L_{\rm p} (E_{\rm p},T_{\rm L})-2.5\log (4\pi F_{\rm p})\nonumber \\
& &	-5\log (3.0857\times 10^{19}),
\label{eq:mu}
\end{eqnarray}
where $F_p$ is the observed energy flux.
We define a chi square function, 
\begin{eqnarray}
&& \chi^{2}(A,B,C)
\propto \sum_{i=1}^{N} \frac{ (\mu_{i}-\mu(E_{p},T_{L},F_{\rm p}))^2}
         {{\sigma^{2}_{\mu_{{\rm local}}, i}+  \sigma^{2}_{\mu_{{\rm GRB}}, i}}
         }, 
\end{eqnarray}
 and 
 \begin{eqnarray}   
&& \sigma^{2}_{\mu_{{\rm GRB}}, i}            
 =        \frac{25}{4}\left[(1 + 2 C) \sigma^2_{\log F_{\rm p,i}} \right. \nonumber \\
&& \left. \hspace{2cm}              + B^2 \sigma^2_{\log E_{\rm p,i}}
                + C^2 \sigma^2_{\log T_{\rm L,i}}
                + \sigma_{\rm int}^2\right],
\label{eq:x_i}
\end{eqnarray}
where $\sigma_{\rm int}$ is the systematic error determined by iteration.

\begin{figure}[ht]
\begin{center}
\includegraphics[width=80mm]{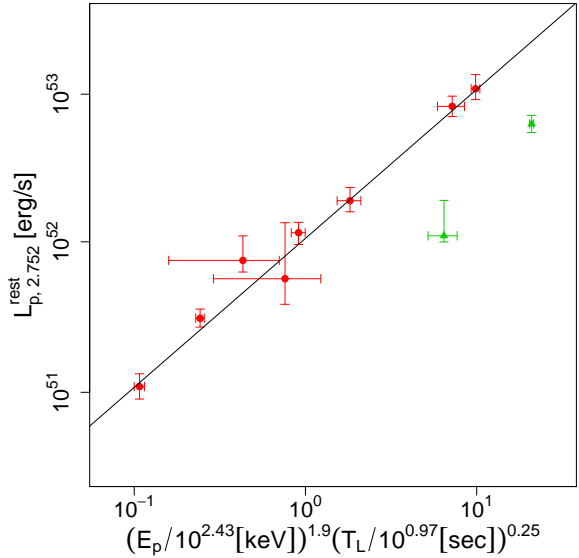}
\vspace{0pt}
\caption{The \tsutsui diagram for small-$ADCL$ events (red filled circles) in $z<1.4$ {\it without}  any assumption of cosmological parameters. 
The solid line is the best fit function  with $\chi^2_{\nu}=0.83$ and $\sigma_{\rm int}=0$. The green triangles are outliers.
 (see \citep{Tsutsui:2012a} for the definition of outliers.)}
\label{fig:FP2}
\end{center}
\end{figure}

Before we minimize the chi square function, we use the robust statistic and outlier detection method which have been developed 
in  Tsutsui et al. \citep{Tsutsui:2011}. 
As a result, we detected two outliers (091003, 080319B)
, and removed them from following chi square analysis.
Performing chi square regression on the remaining samples,  we obtained the best fit parameters and 1-$\sigma$ 
uncertainties summarized in table. \ref{tab:calib}.
The parameters in table. \ref{tab:calib} and outliers of the relation
are consistent with those  in  Tsutsui et al. \citep{Tsutsui:2012a}.
Figure. \ref{fig:FP2} shows the \tsutsui diagram for small-$ADCL$ events in $z<1.4$. 
Two outliers are marked with green triangles and the solid line indicates the best fit function of Eq.(2.1) with values of
A, B and C in table 1.

\begin{table}[htdp]
{
\caption{The best fit values (1-$\sigma$ uncertainties) of the model parameters of the Type I Fundamental Plane (Eq. (2.1)). }
\begin{center}
\begin{tabular}{ccccc}
\hline
A &B & C & $\sigma_{\rm int}$ &$\chi_{\nu}^{2}$ \\
\hline
52.06(0.02) & 1.90(0.05) & 0.25(0.13) & 0  & 0.83  \\
\hline
\end{tabular}
\end{center}
\label{tab:calib}
}
\end{table}
\begin{table}
\begin{center}
\caption{The redshifts, distance modulus of LGRBs  
estimated by local regression method and the ones estimated from the Type I Fundamental Plane.
The first through the seventh column express  name of the GRB,  the redshift, the distance modulus determined by TypeIa
supernovae, the error of $\mu_{\rm local}$, the distance modulus determined by the best fit function of Type I Fundamental Plane
(Eq. (2.1)), the error of $\mu$, the value of $ADCL$, respectively.
Four outliers are separated in the last four rows.
They are the same as the outliers for 
small-$ADCL$ GRBs in Tsutsui et al. \citep{Tsutsui:2012a} }
\begin{tabular}{l|cccccc}
\hline
GRB & $z$ &$\mu_{\rm local}$ & $\sigma_{\mu_{\rm local}}$ & $\mu$ & $\sigma_{\mu}$ & $ADCL$\\
\hline
030329 & 0.168 & 39.48 & 0.04 & 39.61 & 0.17 & 0.13\\
090618 & 0.54 & 42.43 & 0.04 & 42.33 & 0.19 & 0.10\\
050525 & 0.606 & 42.83 & 0.07 & 42.72 & 0.13 & 0.13\\
970228 & 0.695 & 43.15 & 0.05 & 42.7 & 0.72 & 0.14\\
990705 & 0.843 & 43.75 & 0.06 & 43.86 & 0.22 & 0.10\\
091208B & 1.063 & 44.3 & 0.09 & 44.77 & 0.77 & 0.07\\
061007 & 1.261 & 44.73 & 0.06 & 44.78 & 0.16 & 0.14\\
990506 & 1.3 & 44.81 & 0.07 & 44.82 & 0.26 & 0.08\\
070125 & 1.547&-& -& 45.39 & 0.25 & 0.10\\
990123 & 1.6 &-&-& 45.5 & 0.16 & 0.09\\
990510 & 1.619 &-&-& 45.46 & 0.34 & 0.14\\
090902B & 1.822 &-&-& 45.91 & 0.15 & 0.09\\
081121 & 2.512 &-&-& 46.9 & 0.24 & 0.07\\
080721 & 2.602 &-&-& 46.6 & 0.28 & 0.16\\
050401 & 2.9 &-&-& 46.5 & 0.42 & 0.15\\
971214 & 3.418 &-&-& 47.5 & 0.73 & 0.12\\
090323 & 3.57 &-&-& 48.05 & 0.3 & 0.12\\
\hline
091003 & 0.897 & 43.82 & 0.09 & 45.97 & 0.28 & 0.16\\
080319B & 0.937 & 43.82 & 0.08 & 45.36 & 0.17 & 0.14\\
110213A & 1.46 &-&-& 42.9 & 0.38 & 0.09\\
110422A & 1.77 &-&-& 43.51 & 0.16 & 0.10\\
\hline
\end{tabular}
\label{tab1}
\end{center}
\end{table}


\section{Constraint}
\label{sec:constraint}

Applying equation (\ref{eq:mu}) to high redshift LGRBs, we can obtain the extended Hubble diagram with the unprecedented precision. 
We define $\sigma_{\mu_{i}}$ as below, 
\begin{eqnarray}
\sigma^2_{\mu_i}=&&\frac{25}{4}\Big[\sigma_A^2
+(\sigma_B\log{E_{\rm p}}/{\bar E_{\rm p}})^2+(\sigma_C\log{T_{\rm L}}/{\bar T_{\rm L}})^2 \nonumber \\
& &\mbox{ }+(B\sigma_{\log E_{\rm p}})^2+(C\sigma_{\log T_{\rm L}})^2+(1+2C)\sigma_{\log F_{\rm p}}^2\\
& & +\sigma_{\rm int}^2\Big], \nonumber
\label{eq:mu-err}
\end{eqnarray}
Here we neglect the effect of the gravitational lensing because 
M. Oguri and K. Takahashi. \citep{Oguri:2006} discussed the gravitational lensing of GRBs 
and found that the biases are small.

Then we derive constraints on cosmological parameters.
In the $\Lambda$-CDM model, 
the luminosity distance ($d_{L}$) is given as a function of 
the density parameters, $\Omega_{M}$ and $\Omega_{\Lambda}$. 

The chi square function is defined by,
\begin{eqnarray}
\chi^2(\Omega_{M},\Omega_{\Lambda})
&=& \sum_{i=1}^{N} \left[\frac{\mu_i - \mu^{\rm th}(z_i, \Omega_M, \Omega_{\Lambda},)}
       {\sigma_{\mu_i}}\right]^{2}
\end{eqnarray}
where
$\mu^{\rm th}(z_i, \Omega_M, \Omega_{\Lambda})
= 5 \log( d_L^{\rm th}/{\rm Mpc}) + 25$.
We used the robust regression and outlier detection method again.
As a result, we detect two outliers (110213A, 110422A), and removed it from following chi square analysis.
In table \ref{tab1}, we summarize the redshifts, distance modulus of LGRBs  
estimated by local regression method and the ones estimated from the Type I Fundamental Plane
 with their 1-$\sigma$ uncertainties.

\begin{figure}[ht]
\begin{center}
\includegraphics[width=80mm]{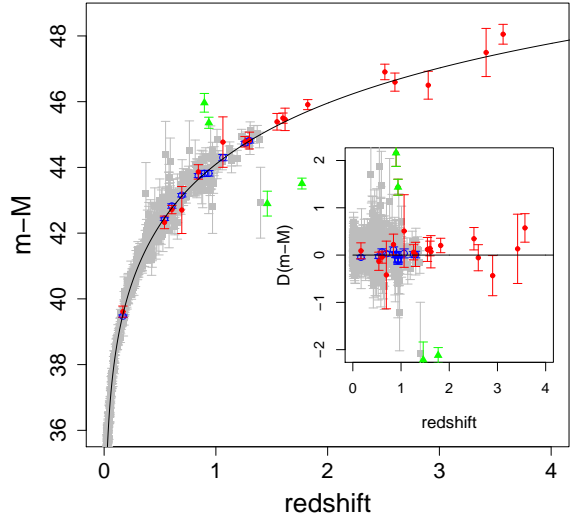}
\vspace{0pt}
\caption{Extended Hubble diagram from the Type I Fundamental Plane.
{\it Gray squares}: the 557 SNe Ia from R. Amanullah et al.\citep{Amanullah:2010}. 
{\it Blue open circlues}: LGRBs at $z<1.4$ estimated from the local regression of SNe Ia data. 
{\it red filled circles}: LGRBs estimated from the Type I Fundamental Plane.
{\it Green triangles}: outliers.
The solid line indicates a theoretical model calculated with ($\Omega_M, \Omega_{\Lambda}$)=(0.30, 0.70).
{\it Inset} : A residual Hubble diagram from the theoretical model.
}
\label{fig:hubble}
\end{center}
\end{figure}

Figure~\ref{fig:hubble} shows an extended Hubble
diagram up to $z = 3.57$.
The gray squares indicate the SNe Ia from R. Amanullah et al. \citep{Amanullah:2010}, 
the blue open circles indicate the low redshift LGRBs at $z<1.4$ estimated from the  local regression 
of SNe Ia data. 
Red filled circles and green triangles indicate LGRBs estimated from the Type I Fundamental Plane, 
and  outliers, respectively. 
The solid line indicates a theoretical model calculated with ($\Omega_M, \Omega_{\Lambda}$)=(0.30,0.70).
As Figure~\ref{fig:hubble} shows, the uncertainties of distance modulus of LGRBs are 
as small as those of individual SNe Ia (gray squares).
The mean error of distance modulus of LGRBs are $\bar \sigma_{\mu}=0.31$, while those of SNe Ia are $\bar \sigma_{\mu}=0.26$.

\begin{figure}[t]
\begin{center}

\includegraphics[width=60mm]{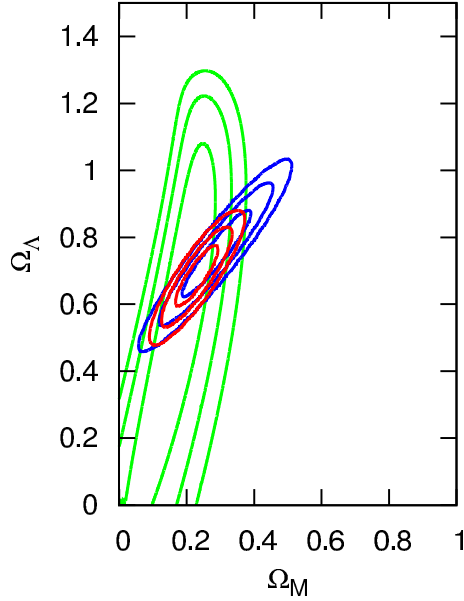}
\caption{Constraints on ($\Omega_M,\Omega_{\Lambda}$) plane from 9 LGRBs with $z>1.4$
 (green), 557 SNe Ia (blue) and 9 LGRBs + 557 SNe Ia (red). The contours correspond
to 68.3\%, 95.4\%, 99.7\% confidence regions.
See also Table~\ref{tab:constraint}.}
\label{fig:contour}
\end{center}
\end{figure}

Performing the chi square analysis on the remaining 9 LGRBs, we obtained  the constraint 
$\Omega_{\rm M}=0.22\pm0.04$ for flat $\Lambda$CDM universe.  
In Figure~\ref{fig:contour}, we show the likelihood contour for non-flat $\Lambda$CDM universe from
9 LGRBs (green), with  the likelihood contours from 
557 SNe Ia (blue) and 9 LGRBs + 557 SNe Ia (red),
and the best-fit values with 1-$\sigma$ errors are shown in table~\ref{tab:constraint}.
Figure 4 shows that adding only 9 high redshift LGRBs ($z>1.4$) to 557 SNe Ia ($z<1.4$)
significantly improves the constraint on $\Omega_{\rm M}$--$\Omega_{\rm \Lambda}$ plane.
 
\begin{table}
\begin{center}
\begin{tabular}[htb]{|c|c|c|c|}
\hline
& $\Omega_M$ & $\Omega_{\Lambda}$ & $\chi^2$/d.o.f  \\
\hline
9 LGRBs (flat) & $0.22\pm0.04$ & - & 5.09/7 \\

\hline
557 SNe Ia (non-flat) & $0.29\pm0.10$ & $0.76\pm0.13$ & 542.1/555 \\
\hline
Combined (non-flat) & $0.23\pm0.06$ & $0.68	\pm0.08$ & 548.3/564 \\ 
\hline
\end{tabular}
\caption{Constraints on ($\Omega_M,\Omega_{\Lambda}$) for flat 
universe from 9 LGRBs and for non-flat 
 universe  from 557 SNe Ia and  9 LGRBs + 557 SNe Ia with their reduced chi squares.
 All the quoted errors are at the 1-$\sigma$ confidence level.}
\label{tab:constraint}
\end{center}
\end{table}

\section{Discussion}
\label{sec:discussion}
In this paper we extended the Hubble diagram with the Type I Fundamental Plane of LGRBs. 
Because of the tightness of the relation, our distance measurements are much more precise than 
those of previous works. 
Although the number of LGRBs are much smaller than that of the previous works \citep{Schaefer:2007, Kodama:2008,Liang:2008,
Amati:2008,Tsutsui:2009a,Cardone:2009,Tsutsui:2009b}, our constraint is much stronger. 
This indicates the importance of controlling and removing systematic errors.
Although we used only the Type I Fundamental Plane to measure distance modulus, 
in principle, we may also use the Type II Fundamental Plane as an independent distance indicator.
However, because there is still a small number of long tailed events,  it is difficult to calibrate the relation 
with our outlier rejection technique. 

The constraints from 9 high redshift GRBs are consistent with those from other probes, e.g. 
Cosmic Microwave Background (CMB), Baryon Acoustic Oscillation (BAO)\citep{Komatsu:2011,Eisenstein:2005}.
We should stress here that this consistency is not trivial. Since the redshift range of GRB ($1.4 < z <  3.5$) is different from
TypeIa supernovae ($z < 1.7$), BAO ($z\sim 0.35$) and CMB ($z=z_{rec}$), they might show different cosmological parameters if
the dark energy is strongly time dependent. Therefore our results suggest that time dependence of the dark energy is not so strong
even if it exists.

The purpose of this paper is to show the potential high ability of Type I Fundamental Plane of the LGRBs 
as a distance indicator. Combining our data with other cosmological probes is the next step. 
 A different direction is to test more general equation of state than $\Lambda$CDM universe.
 We are now trying to combine our data with CMB and BAO 
as well as more general equation of state to constrain  the property of the dark energy.

\section*{Acknowledgments}
This work is supported in part by the Grant-in-Aid for Grant-in-Aid for Young Scientists (B) 
from the Japan Society for Promotion of Science (JSPS), No.24740116(RT),  by the Grant-in-Aid from the 
Ministry of Education, Culture, Sports, Science and Technology
(MEXT) of Japan, No.23540305 (TN), No.20674002 (DY), No.23740179 (KT),
and by the Grant-in-Aid for the global COE program 
{\it The Next Generation of Physics, Spun from Universality 
and Emergence} at Kyoto University.


\providecommand{\href}[2]{#2}\begingroup\raggedright\endgroup

\end{document}